\documentclass[runningheads]{llncs}
\usepackage[T1]{fontenc}
\usepackage{amsmath}
\usepackage{amssymb}
\usepackage{amsfonts}
\usepackage{amstext}
\usepackage{graphicx}
\usepackage{multirow}%

\begin{document}
\title{Automatic facial axes standardization of 3D fetal ultrasound images}
%
%\titlerunning{Abbreviated paper title}
% If the paper title is too long for the running head, you can set
% an abbreviated paper title here
%
\vspace{-3mm}
\author{Antonia Alomar\inst{1} \and
Ricardo Rubio\inst{2,3}\and Laura Salort \inst{1} \and
 Gerard Albaiges\inst{4} \and
 Antoni Payà\inst{2,3}\and
 Gemma Piella\inst{1}\and
 Federico Sukno\inst{1}}
% %
% % \orcidID{0000-0003-3658-5832}
% % \orcidID{00000-0003-0025-2614} 
% % \orcidID{0000-0001-8690-3996}
% % \orcidID{0000-0002-9036-1349}
% % \orcidID{0000-0001-5236-5819}
% % \orcidID{0000-0002-2029-1576}
\authorrunning{A. Alomar et al.}
% % First names are abbreviated in the running head.
% % If there are more than two authors, 'et al.' is used.
% %
\institute{Department of Information and Communications Technologies, Universitat Pompeu Fabra, 122-140 Tànger, Barcelona, Spain \and
Department of Obstetrics and Gynecology, Hospital del Mar, 25-29 Passeig Marítim, Barcelona,Spain \and Department of Medicine and Life Sciences, Universitat Pompeu Fabra, 88 Doctor Aiguader,Barcelona, Spain \and Fetal Medicine Unit, Obstetrics Service, Department of Obstetrics, Gynecology and Reproductive Medicine, University Hospital Quirón Dexeus, Barcelona, Spain}
\maketitle              % typeset the header of the contribution
\vspace{-5mm}
\begin{abstract}
Craniofacial anomalies indicate early developmental disturbances and are usually linked to many genetic syndromes. Early diagnosis is critical, yet ultrasound (US) examinations often fail to identify these features. This study presents an AI-driven tool to assist clinicians in standardizing fetal facial axes/planes in 3D US, reducing sonographer workload and facilitating the facial evaluation. Our network, structured into three blocks—feature extractor, rotation and translation regression, and spatial transformer—processes three orthogonal 2D slices to estimate the necessary transformations for standardizing the facial planes in the 3D US. These transformations are applied to the original 3D US using a differentiable module (the spatial transformer block), yielding a standardized 3D US and the corresponding 2D facial standard planes. The dataset used consists of 1180 fetal facial 3D US images acquired between weeks 20 and 35 of gestation. Results show that our network considerably reduces inter-observer rotation variability in the test set, with a mean geodesic angle difference of 14.12$^{\circ}$ ± 18.27$^{\circ}$ and an Euclidean angle error of 7.45$^{\circ}$ ± 14.88$^{\circ}$. These findings demonstrate the network's ability to effectively standardize facial axes, crucial for consistent fetal facial assessments. In conclusion, the proposed network demonstrates potential for improving the consistency and accuracy of fetal facial assessments in clinical settings, facilitating early evaluation of craniofacial anomalies.
\end{abstract}
\section{Introduction}
Craniofacial anomalies serve as indicators of developmental disturbances at early stages of life, encompassing a wide range of heterogeneous conditions associated with many genetic syndromes \cite{Junaid2022,Bartzela2017}. Estimates suggest that up to 40\% of genetic syndromes produce alterations in the normal morphology of the face and the head. Although these associations have predominantly been identified in adult populations, there is increasing interest in early assessment \cite{Tavares2022}. Consequently, diagnostic efforts are moving towards prenatal and postnatal stages \cite{Chen2022}. %,Liu2024.

To evaluate the fetal development, 2D ultrasound (US) imaging is the standard procedure. Unfortunately, dysmorphology features are hard to identify in this way, due to the noisy nature of fetal US (low signal-to-noise ratio, fetal or probe movements, fetal position, and limbs in front of the face) \cite{Conner2014}. Currently, 3D/4D US serves as a complement to 2D US. They prove to be particularly useful in diagnosing various fetal anomalies, especially those involving facial abnormalities, neural tube defects, and skeletal anomalies \cite{Dyson2000,Merz2017}. 

In this context, acquiring an US standard plane (SP) is crucial for performing an accurate fetal diagnosis, as the SP is used to measure and analyse biomarkers and abnormal features \cite{Salomon2019,Sarris2011}. 3D US has the advantage of capturing multi-view planes allowing sonographers to manually select SPs from 3D US images or videos during prenatal exams. While this process is essential, it is also time-consuming and observer-dependent. This manual selection can be laborious and biased due to the extensive search space, the sonographer experience, and the variability of the fetus orientation \cite{Nerea_2024,Sarris2012}. 

In this study, we aim to reduce the sonographer's workload while enhancing the accuracy and interpretability of fetal facial SP detection. We propose an AI-driven tool designed to assist clinicians in standardizing the facial axes/planes in the 3D US. It aims to minimize variability across planes detection while mitigating the effects of clinician subjectivity in selecting accurate SPs for fetal facial assessment. Standardizing the fetal facial axes intends to facilitate the evaluation of facial biomarkers and biometric measurements to perform facial assessment. The proposed method consists in regressing the transformation necessary to standardize the sagittal, coronal and axial fetal facial axes, taking as input 3 orthogonal planes centered at the middle of the 3D US image. The novelty lies in that instead of combining the regression model with another task, such as the classification of the planes, we add a differentiable block that incorporates the image loss between estimated and ground truth (GT) planes as part of the minimization strategy. This helps the network learn the structures that should be present in the SPs. Additionally, the proposed algorithm offers the advantage of low computational cost and easy integration into in the echographer or in 3D Slicer as a built-in feature. 
%- not computationally expensive to be feasiblet incorponate it either in 3D ultraspund r in 3D slicer as a post prosses toool
%- estimate the translation necessary to estimate standardize the 3D US --> independent from the original image resolution so no information is lost
\vspace{-3mm}
\section{Related Work}
Several methods have been proposed for automatically detecting 2D SPs in US images using deep learning. Usually, this task has been approached as an image classification problem, using convolutional neural networks (CNNs) or recurrent neural networks (RNNs) \cite{Chen_2017,Baumgartner2017,Lei2015,Zhen2023}. However, these methods only determine whether the acquired 2D slices are SPs, but do not inform on what correction shall be applied to them in case they are not SPs.

Another common strategy consists of regressing the plane parameters or transformation matrices to achieve the SP in the 3D US volume. For example, Feng et al.\cite{Feng_2009} proposed a constrained marginal space learning method that combines both 2D and 3D information for fetal face detection in 3D US. Nie et al. \cite{Nie2017} introduced a deep belief network combined with a detection algorithm to provide a prior structural knowledge to the network. Li et al. \cite{Li2018} presented an iterative transformation network to detect SPs in 3D fetal US using a CNN that performs both plane classification and regression to estimate the transformation parameters. Di Vece et al. \cite{Di_Vece_2024} improved the previous results obtained estimating the six-dimensional pose of arbitrary oriented US plane of the fetal brain with respect to a template normalized frame using a CNN regression network. Recently, reinforcement learning (RL) has shown great potential in addressing SP localization as a regression task \cite{Dou2019,Huang2020,Zou2022,Li2021}. Although RL approaches have achieved high performance, several issues remain to be addressed, such as the reliance of current studies on initial registration to ensure data orientation consistency, which can easily fail if the pre-registration process is unsuccessful. Moreover, unlike parameters regression models using a CNN, RL models simplify the problem of regressing the transformation parameters by considering a discrete action space and, in consequence, limiting the transformation that can be applied. 
To avoid dependence on pre-registration and ensure no limitations on the transformations that can be performed, we choose a classical yet effective parameters regression approach using a CNN. To help the network learn the structures that should be present in the SPs, we add a differentiable block that incorporates the image loss between estimated and GT planes as part of the minimization strategy. As a result, the number of parameters of the network is significantly reduced because no classification blocks are used.

%Due to the high similarity between the standard and nonstandard planes, high intra-class variations of standard plane caused by various gestational ages, different fetal postures, various scanning orientations, and the presence of speckles and artifacts in US images, automatically recognizing standard planes remains challenging \cite{Skelton2021}. For this reason, we aim to standardize the facial axes planes by regressing the transformation necessary to transforms the input 3D US such as the resulting volume have the facial sagittal, coronal and axial SPs centered in the middle of the transformed US. These will facilitate clinicians the removal of the scanning orientation of the fetal face to facilitate its evaluation. The proposed network take as input the 3 orthogonal planes centered at the middle of the 3D US, and estimates the transformation matrix that represent the change of bases to the standardized fetal face axes.
\vspace{-3mm}
\section{Method}
\label{sec:methods}
\begin{figure}
\includegraphics[width=\textwidth]{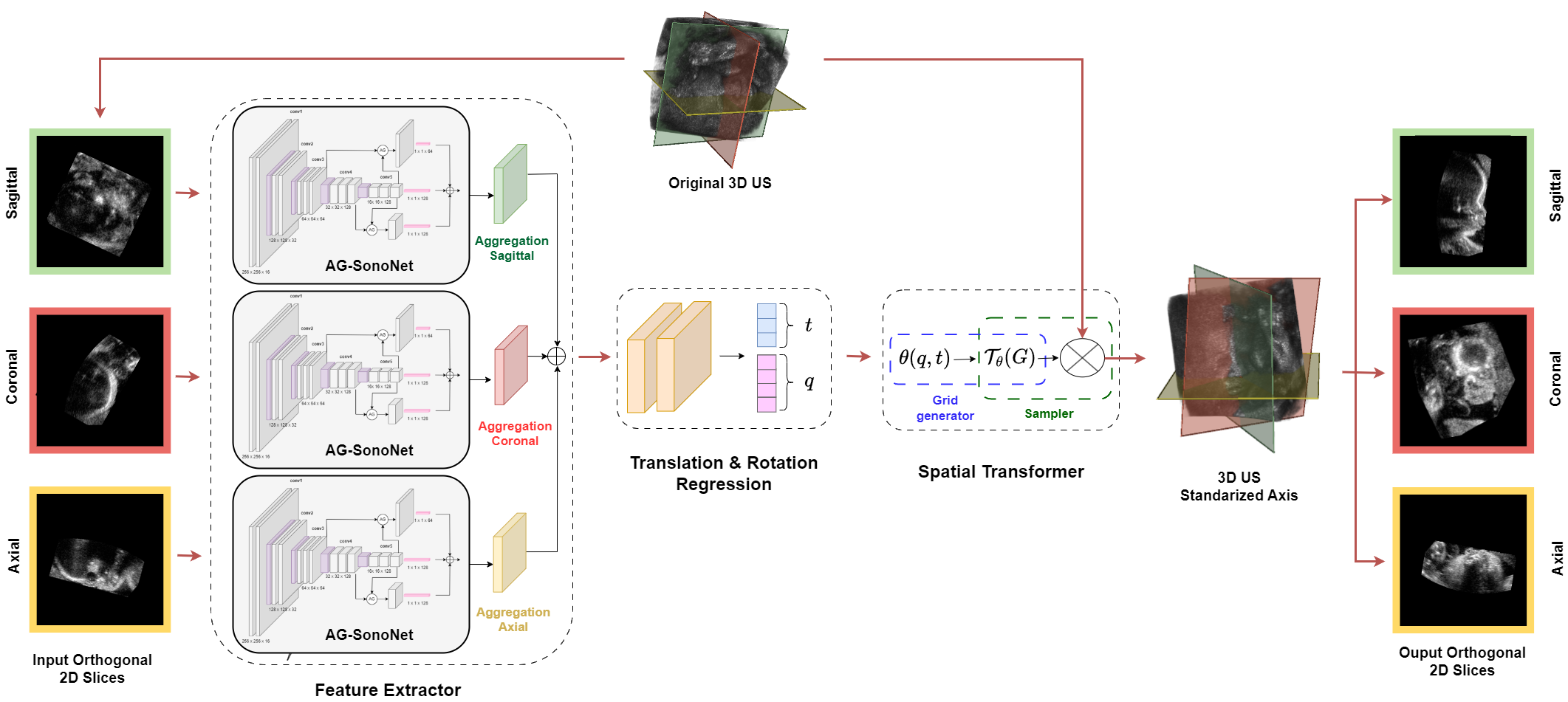}
\caption{Proposed architecture. The network is divided in three blocks: feature extractor, rotation and translation regression, and the spatial transformer block. The two first blocks extract the features and estimate the transformation necessary to obtain the standard facial axes/planes taking as input three orthogonal slices in arbitrary position. Then, the differentiable spatial transformer block applies the estimated transformation to the original 3D US to obtain the standardized 3D US and the 2D facial SPs.} \label{fig:overview}
\end{figure}
\subsubsection{Data \& Pre-processing:}
The dataset used consists of 1180 fetal facial 3D US images acquired between week 20 and 35 of gestation (26.56 $\pm$ 2.72) using a Voluson E8 RSA (BT-20) with a convex probe (4D-RAB6-D, 2–8 MHz) at two hospitals in Barcelona (Hospital de Mar and Hospital Universitari Dexeus) according to their Ethical Research Committee and the current legislation (Organic Law 15/1999). The study population comprises subjects from low-risk pregnancies, meaning without any pathology, or known family cases of craniofacial or syndromic pathologies, which were all carried to term. The data is divided into training, validation, and test sets, with 72\%, 12\%, and 16\% of the data allocated to each set, respectively. To facilitate the use of deep learning, it is essential to standardize the input image size across the training, validation, and test sets. We implemented this through two steps: 1) down-sampling the 3D US image by a factor of two, to reduce the computational cost of the network; 2) symmetric zero-padding to the center of the 3D US to achieve a size of $U \in \mathbb{R}^{C\times H \times W \times D}$ where $H,W,D = 256$ are the height, width, and depth dimension and $C= 1$. The latter is performed to ensure that no information is cropped-out during rotation and translation. The 2D initial planes are defined by $I_{0} = [I_s, I_c, I_a]$ with $I_s = U(1,\frac{H}{2}, : , :)$, $I_c = U(1,;, \frac{W}{2} , :) $ and $I_a = U(1,;, ; , \frac{D}{2})$, corresponding to the sagittal, coronal, and axial planes, respectively.
%we down-sample the 3D US images by a factor of 2. Moreover, to ensure that no information is lost when rotating and translating, the image center is computed and zeros are added symmetrically to achieve a size of $U \in \mathbb{R}^{H \times W \times D\times C}$ where $W,H,D = 256$ are the height, weigh and depth dimension and $C= 1$ is the channel dimension.
\vspace{-5mm}
\subsubsection{Ground Truth Standard Planes:}
The facial GT SPs we are interested in locating are the axial, coronal and sagittal planes that define the canonical axes of the fetal face. They are obtained by minimizing the 3 orthogonal planes defined by 23 anatomical landmarks located by expert clinicians in the 3D US (see Appendix Fig. 1) and following the recommendations from the international 3D focus group \cite{Merz2012}. We constrain the planes' normal vectors $\overrightarrow{n_a} ,\overrightarrow{n_c}$, and $\overrightarrow{n_s}$ to be orthonormal. The center of the planes $\overrightarrow{c} = (c_x, c_y, c_z)^{T}$ is defined as the intersection point of the 3 planes. The GT is obtained using custom code in 3D Slicer. The extracted normal vectors of the 3 orthogonal planes are used to compute the rotation matrix $R_{gt} \in \mathbb{R}^{3\times3}$ needed to standardize the image axes to the estimated facial SPs. The GT rotation can be written as the change-of-basis matrix $S_{\mathbb{B}_1 \rightarrow \mathbb{B}_2 }$ from $\mathbb{B}_1 \rightarrow \mathbb{B}_2 $. In our case, $\mathbb{B}_1$ coordinates correspond to the canonical basis and $\mathbb{B}_2$ coordinates are the estimated normal vectors. Thus,
\begin{equation}\label{eq:rotation_gt}
    R_{gt}= S_{\mathbb{B}_1 \rightarrow \mathbb{B}_2 } =  \begin{pmatrix}
        \overrightarrow{n_s}^T \\\overrightarrow{n_c}^T \\ \overrightarrow{n_a}^T 
    \end{pmatrix} = \begin{bmatrix}
    r_{11} & r_{12} & r_{13} \\
    r_{21} & r_{22} & r_{23} \\
    r_{31} & r_{32} & r_{33} \\
\end{bmatrix}
\end{equation}
The rotation regression is performed in terms of quaternion representation for compactness, i.e., $\overrightarrow{q_{gt}} = (q_0, q_1, q_2, q_3)^{T}$. The conversion from rotation matrix to quaternion follows
\vspace{-3mm}
\begin{equation}
    \overrightarrow{q_{gt}} = \begin{pmatrix}
    q_0\\ q_1\\ q_2 \\ q_3\\ \end{pmatrix}  \\ =\begin{bmatrix}
    \frac{1}{2} \sqrt{1 + r_{11} - r_{22} - r_{33}}\\
    \frac{r_{12} + r_{21}}{4 q_0}\\
    \frac{r_{13} + r_{31}}{4 q_0}\\
    \frac{r_{23} - r_{32}}{4 q_0}\\
\end{bmatrix}  
\end{equation}
The intersection/center of the planes is the GT translation $\overrightarrow{t_{gt}} = \overrightarrow{c} \in \mathbb{R}^{3}$. Then, the GT transformation matrix is 
\vspace{-3mm}
\begin{equation}
    \theta_{gt} = \begin{pmatrix}
    R_{gt} ,& \overrightarrow{t_{gt}} \\ 
\end{pmatrix} =\begin{bmatrix}
    r_{11} & r_{12} & r_{13} & t_x\\
    r_{21} & r_{22} & r_{23} & t_y\\
    r_{31} & r_{32} & r_{33} & t_z\\
\end{bmatrix}  \in \mathbb{R}^{3\times4}
\end{equation} 
To ensure compatibility with the spatial transformer block, the translation is expressed in image relative size. Thus, each component of $\overrightarrow{t_{gt}}$ is in the range $[-1,1]$. The 2D GT sagittal, coronal and axial SPs are obtained as $I_{gt} = [I_s, I_c, I_a]$ where $I_s = V_{gt}(1,\frac{H}{2}, : , :)$, $I_c =  V_{gt}(1,;, \frac{W}{2} , :) $ and $I_a =  V_{gt}(1,;, ; , \frac{D}{2})$, and $ V_{gt}$ is the transformed US using $\theta_{gt}$.
\vspace{-5mm}
\subsubsection{Feature Extractor Block:}
The inputs of the feature extractor block ($I_0 \in \mathbb{R}^{H\times W \times 3} $) are the 3 orthogonal planes located at the center of the 3D US image (sagittal, coronal and axial plane). Each branch uses the AG-SonoNet \cite{Schlemper_2019} as the backbone feature extractor, with the weights shared among the three branches. Then, each view has a specialized aggregation block that adds the attention information from multiple layers of the network to extract specialized features from each view. This information is concatenated and fed to the translation and rotation regression block. 
\vspace{-5mm}
\subsubsection{Translation \& Rotation Regression Block:} It consists of two fully connected layers that convert the extracted features from the three orthogonal planes into the translation and rotation necessary to achieve the standardized axes/planes. The output is the regression vector $\overrightarrow{z} \in \mathbb{R}^7$. The first three positions correspond to the translation vector $\overrightarrow{t_{es}} \in \mathbb{R}^3$ where $\overrightarrow{t_{es}}= (t_x,t_y,t_z)^{T}$ with each component being in the range $[-h_{max},h_{max}]$. The remaining 4 positions correspond to the quaternion representation of the rotation matrix $\overrightarrow{q_{es}}= (q_0,q_1,q_2,q_3)^{T}$ with each component being in the range $[-1,1]$. To represent a valid rotation, $||\overrightarrow{q_{es}}||$ needs to be 1. To ensure that this condition is satisfied, a normalization layer was added after the last fully connected layer. Given $\overrightarrow{q_{es}}$, the estimated rotation can be expressed as:
\begin{equation}
R_{es} = \begin{bmatrix}
1 - 2(q_2^2 + q_3^2) & 2(q_1q_2 - q_0q_3) & 2(q_1q_3 + q_0q_2) \\
2(q_1q_2 + q_0q_3) & 1 - 2(q_1^2 + q_3^2) & 2(q_2q_3 - q_0q_1) \\
2(q_1q_3 - q_0q_2) & 2(q_2q_3 + q_0q_1) & 1 - 2(q_1^2 + q_2^2)
\end{bmatrix} = \begin{bmatrix}
    r_{11} & r_{12} & r_{13} \\
    r_{21} & r_{22} & r_{23} \\
    r_{31} & r_{32} & r_{33} \\
\end{bmatrix}
\end{equation}
\vspace{-7mm}
\subsubsection{Spatial Transformer Block:} It is a differentiable module capable of applying spatial transformations to the original 3D US image $U \in R^{H\times W \times D \times C}$, resulting in a new standardized 3D US image $V \in R^{H \times W \times D \times C}$ and facial 2D SPs. The spatial transformer uses a differentiable 3D bi-linear sampling as defined in \cite{Jaderberg_2015}. Each output value for pixel i can be written as
\begin{equation}
    V_i = \sum^H _n \sum^W_m \sum^D_l U^c_{n,m,l} \max(0,1-|x^s_i -m|) \max(0,1-|y^s_i -n|) \max(0,1-|z^s_i -l|)
\end{equation}
where $(x_i^{inp}, y_i^{inp}, z_i^{inp})$ are the input coordinates that define the sampling points in the original 3D US $U$ and $U_{nml}$ is the value of $U$ at location $(n,m,l)$. The output coordinates $(x_i^{out}, y_i^{out}, z_i^{out})$ are defined to lie on a regular grid $G = {G_i}$ of pixels $G_i = (x^{out}_i , y^{out}_i , z^{out}_i)$ and can be obtained by a 3D affine transformation: 
\vspace{-3mm}
%and the estimated $\theta_{es}$ transform to transform the original 3D Us to obtain the standardized 3D US facial axes. 
\begin{equation}
\begin{pmatrix}
    x^{inp}_i \\
    y^{inp}_i \\
    z^{inp}_i \\
\end{pmatrix} = \mathcal{T}_\theta(G) = \theta(\overrightarrow{q},\overrightarrow{t}) \begin{pmatrix}
    x^{out}_i \\
    y^{out}_i \\
    z^{out}_i \\
    1\\
\end{pmatrix} =\begin{bmatrix}
    r_{11} & r_{12} & r_{13} & t_x\\
    r_{21} & r_{22} & r_{23} & t_y\\
    r_{31} & r_{32} & r_{33} & t_z\\
\end{bmatrix}  \begin{pmatrix}
    x^{out}_i \\
    y^{out}_i \\
    z^{out}_i \\
    1\\
\end{pmatrix}
\end{equation}
$\theta (\overrightarrow{q},\overrightarrow{t})$ is the 3D transformation matrix estimated. We use height, weight and depth normalized coordinates, such that $x_i,y_i,z_i \in [-1,1]$.  The 2D estimated SPs are obtained  as $I_{es} = [I_s, I_c, I_a]$ where $I_s = V(1,\frac{H}{2}, : , :)$, $I_c = V(1,;, \frac{W}{2} , :) $ and $I_a = V(1,;, ; , \frac{D}{2})$.

%GT rotation and translation are define in terms of the relative image size. Then, at inference the estimated rotation and translation can be used on the the original image size obtaining the facial axes standardized 3D US.
\vspace{-3mm}
\subsubsection{Cumulative Transformations \& Initialization:}
At initialization time ($it=0$), $R_0$ is a random rotation defined by the Euclidean angles $\alpha_x, \alpha_y, \alpha_z \in [-20,20]$ degrees and $\overrightarrow{t_0}  =(t_x,t_y,t_z)^{T}$ is random translation with each component being in the range $[-0.05, 0.05]$. To preserve image quality, we accumulate the transformations and perform a unique transformation to the input 3D image. If multiple steps of the network are applied, we define the transformations at step $it$ as $R_{es}^{it} = R_{es}^{it-1} R_{es}$ and $\overrightarrow{t_{es}^{it}}= \overrightarrow{t_{es}^{it-1}} + \overrightarrow{t_{es}}$, whereas $R_{gt}^{it} = R_{es}^{-1} R_{gt}^{it-1}$ and $\overrightarrow{t_{gt}^{it}} = - (R_{es}^{it-1})^{-1} \overrightarrow{t_{es}} + (R_{es}^{it-1})^{-1} \overrightarrow{t_{gt}}$. Here, $R_{es}$, $t_{es}$ correspond to the rotation and translation estimated by the CNN at the current iteration, whereas $R_{es}^{it}$, $t_{es}^{it}$ denote the accumulated rotation and translation at iteration it. We found that 3 iterations are enough to improve performance by refining the SP estimates. However, there was no need to train the network in an iterative way.
\vspace{-3mm}
\subsubsection{Network Loss:} It
is defined as a combination of the mean absolute error (MAE) between the GT and estimated translation, the relative angle (SO3) between the GT and estimated rotation and the image loss between computed as the Frobenius norm of the difference between the GT and estimated SPs:

%The distance between two rotations RA ∈ SO(3) and RB ∈ SO(3) can be obtained asthe rotation angle θAB corresponding to the relative rotation RAB = RT ARB:
\vspace{-5mm}
\begin{equation}
   \mathcal{L} =   \beta ||\overrightarrow{t_{es}} - \overrightarrow{t_{gt}}||_1 + \gamma \quad acos(0.5 * (Tr(R_{es} (R_{gt})^T)-1))+ ||I_{gt} - I_{es}||_1
\end{equation}
where $Tr$ correspond to the trace, $\beta$ and $\gamma$ are the translation and rotation weights in the loss, $I_{gt}$ and $I_{es}$ are the 2D GT and estimated slices corresponding to the fetal facial sagittal, coronal and axial SPs.
%Riemannian/Geodesic/Angle metric 
%$$d_R (R1,R2) = \frac{1}{\sqrt{2}} ||log (R^-1_1 R_2)||_F$$

\section{Experiments}

The proposed method is compared to the inter-observer variability obtained from 3 different observers placing landmarks on the 3D US images as described in González-Aranceta et al. \cite{Nerea_2024}, which are then used for estimating the planes as described in Section \ref{sec:methods}. Next, we compare to the state of the art method proposed by Li et al. \cite{Li2018}. For a fair comparison, the network used is their M1 baseline model with the addition of the differential spatial transformer model to be able to train with the image loss. The only task learned is the regression of $\overrightarrow{t},\overrightarrow{q}$using as input 3 orthogonal planes. The predicted SPs/axes are evaluated in the test set against the GT using the distance between the GT and the estimated translation, and the rotation angles between the GT and the estimated rotation. The image similarity of the planes is also measured using the peak to noise-ratio (PSNR) and structural similarity index (SSIM).

\section{Results \& Discussion}
\vspace{-4mm}
\begin{table}
\caption{Quantitative test set comparison of the 3D standard facial axes standardization. Metrics evaluated are the geodesic angle difference (SO3), mean Euclidean angle error (EA), mean absolute translation (Trans) error, SSIM, and PSNR for estimated 2D standard planes (SP). The arrows indicate if higher/lower values are better.}\label{tab:com_1}
\centering
\begin{tabular}{l|l|l|l|l|l}
\hline
\textbf{Method} &  \textbf{SO3 ($^{\circ}$) } $\downarrow$  & \textbf{EA($^{\circ}$) } $\downarrow$  & \textbf{Trans(mm)} $\downarrow$ & \textbf{SSIM(\%)}$\uparrow$ & \textbf{PSNR(dB)}  $\uparrow$ \\
\hline
Inter-observer &  21.91$\pm$34.38 & 10.35$\pm$19.33 & 3.31$\pm$2.46& 0.73$\pm$0.10 & 23.46$\pm$7.90\\
\hline
Li et al. \cite{Li2018} & 30.33$\pm$36.34 & 17.32$\pm$30.27& 13.30$\pm$12.99 & \textbf{0.62$\pm$0.08}& \textbf{17.90$\pm$2.87}\\
\hline
Ours & \textbf{14.12$\pm$18.27} & \textbf{7.45$\pm$14.88} & \textbf{12.89$\pm$6.07} & 0.61$\pm$0.07 &  16.98$\pm$1.62 \\
\hline
\end{tabular}
\end{table}
\vspace{-3mm}
Table \ref{tab:com_1} summarizes the results obtained. The proposed method outperforms the state of the art method from Li et al. \cite{Li2018} and even challenges inter-observer variability, producing smaller angular errors although larger translation errors.
\vspace{-3mm}
\begin{table}
\centering
\caption{Quantitative test set comparison per plane/axes. The rotation and translation are are separated in plane/axes components.} \label{tb: com2} 
\begin{tabular}{l|l|l|l|l}
\hline
\textbf{Metric} &\textbf{Model} & \textbf{Sagittal} & \textbf{Axial} &\textbf{Coronal}\\
\hline
\multirow{3}{*}{Translation (mm)} & Inter-observer & 1.60 $\pm$ 1.74 & 1.94 $\pm$ 1.84 & 1.51 $\pm$ 1.40\\
\cline{2-5}
& Li et al. \cite{Li2018} & \textbf{6.42 $\pm$ 8.89} & 5.97$\pm$ 8.41& 7.14 $\pm$ 8.25\\
\cline{2-5}
& Ours & 7.26 $\pm$ 6.57 & \textbf{5.78 $\pm$ 4.20} & \textbf{5.76 $\pm$ 4.78}\\
\hline
\multirow{3}{*}{Rotation ($^{\circ}$)}& Inter-observer & 11.35 $\pm$ 21.66 &  7.08 $\pm$ 9.60 &  12.63 $\pm$ 26.73\\
\cline{2-5}
& Li et al. \cite{Li2018} & 19.29 $\pm$ 33.21& 10.88 $\pm$ 11.42 & 21.81 $\pm$ 34.12\\
\cline{2-5}
& Ours & \textbf{8.70 $\pm$ 19.69}& 
 \textbf{6.14 $\pm$ 6.73}& \textbf{7.49 $\pm$ 18.88}\\
\hline
\end{tabular}%\vspace*{.2mm}
\end{table}

\begin{figure}
\centering
\includegraphics[width=12cm]{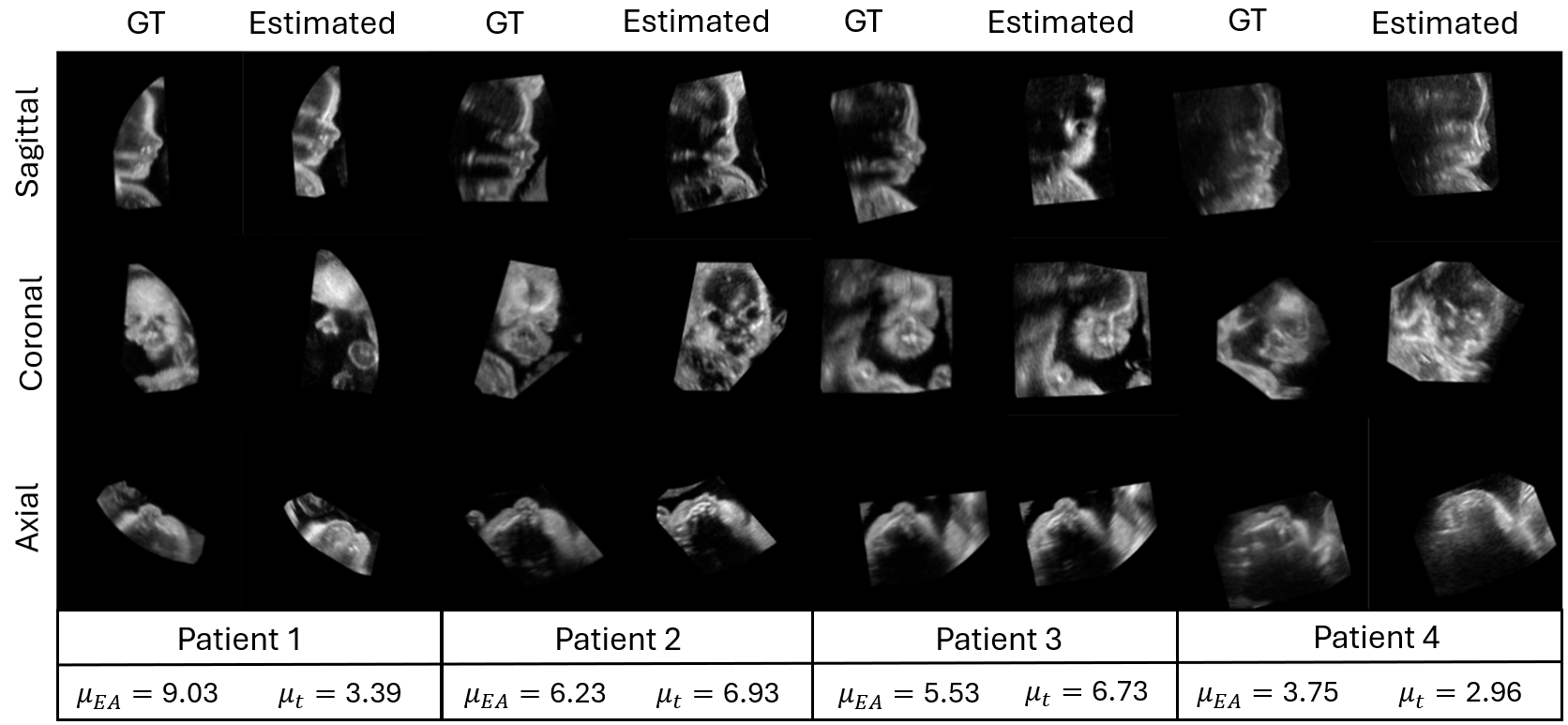}
\caption{Qualitative test set examples. Four examples of the estimated 2D planes using the proposed method, the mean Euclidean angle ($\mu_{EA}$) in degrees and the mean translation ($\mu_{t}$) error in mm obtained per plane compared to the GT. } \label{fig:examples}
\end{figure}

\vspace{-3mm}
Table \ref{tb: com2} shows the translation and rotation Euclidean angles error obtained per SP/axes. It highlights that the coronal and sagittal planes are more challenging to locate in terms of rotation than the axial plane. Our approach reduces the inter-observer rotation error of the sagittal, axial and coronal planes. The translation error obtained is around 6mm per plane/axes, higher than the inter-observer error. Fig. \ref{fig:examples} shows some qualitative examples and comparisons between the GT 2D planes and the estimated by the proposed method. Despite a translation error per plane of approximately 6 mm in patient 2 and 3, the estimated plane closely approximates the GT plane.

% and the estimated by the Li et al. \cite{Li2018} 
Thus, the proposed network correctly learns the plane angles but its translation error is higher than the inter-observer error. This could be due to the high variability defining the planes localization, as multiple slices could closely resemble each other \cite{Skelton2021}. However, it could also be that the most informative planes for the network are not the ones defined as GT planes and it sacrifices translation accuracy for rotation accuracy. Although the translation errors obtained with the proposed network are larger than the inter-observer, the mean translation error per plane is around 6 mm. Moreover, rotation can be regarded more important than translation, as the aim is to standardize the facial axes. This allows the sonographer to examine the standardized fetal 3D facial US, where the fetal pose/US probe rotations are removed, facilitating the evaluation of the fetal face. Furthermore, the proposed network is able to reduce the rotation inter-observer variability. This can be highly beneficial to homogenize the facial analysis evaluation criteria and to reduce the reliance on clinician expertise while reducing the time burden of manually locating the planes.  
\vspace{-3mm}
\section{Conclusions}
We propose a network that estimates the transformation necessary to obtain the standard sagittal, coronal and axial facial US axes taking as input three 2D orthogonal planes. Evaluation on 184 US volumes shows that the network correctly standardizes the US 3D axes while reducing the rotation variability across observers. The method has the potential to be applied easily in a clinical setting due to its low computational cost. The standardization of the facial 3D US aims to facilitate the analysis of the facial biometric measurements to assess the presence of craniofacial abnormalities.
 %- Method invariant to image size. 
% - canonical facil axes standaritzation of the 3D US.
 %- low computational cost.
% - facilitates the analysis or performance of biometric measurements of the face. 
 
\begin{credits}
\subsubsection{\ackname} This work was partly supported by grants PID2020-114083GB-I00 and PRE2021-097544 funded by MICIU/AEI/10.13039/501100011033/ and under the ICREA Academia programme.
\subsubsection{\discintname}
None of the authors have any competing interests.
\end{credits}
%
% ---- Bibliography ----
%
% BibTeX users should specify bibliography style 'splncs04'.
% References will then be sorted and formatted in the correct style.
%
\bibliographystyle{splncs03}
\bibliography{Refences}
%
%\appendix
%\input{appendix}
\end{document}